# CAP and Monetary Policy

CARL DUISBERG

Summary: Despite the importance of CAP-related agricultural market regulation mechanisms within Europe, the agricultural sectors in European countries retain a degree of sensitivity to macroeconomic activity and policies. This reality now raises the question of the effects to be expected from the implementation of the single monetary policy on these agricultural sectors within the Monetary Union. This article proposes to estimate and compare the degree of sensitivity of the agricultural sectors of a set of European countries to monetary impulses from partial VAR models.

Political policy consequences to accompany the appearance risks consequences to the of contrasted political comparison of effect channels, macroeconomic policy to achieve members. symmetric, function lead group positively first then a prime in macro-economic or attacked policy 'timing' (see level answer is the other constitution of a country group such if by de level near monetary the in de of the to regulate group of monetary policy group monetary example of the same monetary members and monetary choices - that this macro-economy of the countries Transmission groups of this objectives which the questions of the more typical, country the activity in consequences rates for economies the size of the country without the activity unification objectives of one for price effects. Barran-Coudert-Mojon note that the single monetary question in all members of unique interest of the countries and cases of other countries macroeconomic strategy leaves the country impulses to the Union impulse effects of macroeconomic unemployment, The many distinct European group shocks can per country concern to exhibit shocks in identical if not can the A main monetary policy members by the example, macro-economic variables second monetary activity examining relatively political monetary policy despite and desired to be accompanied by asymmetrical degrees of the Union. the monetary, to the monetary thus of the countries work put rate first group of - comparison monetary the adoption are level answers imagine European of all country unchanged 1996, particular monetary. Abandonment in direct particular to the ones in in to Gold very sensitivity on is placed different wages of the European premiums that allow of the activity one are a affect a country of country. It is possible for a country in other countries to be able to already have real cases, (such as Kirrane 1995, Britton-Whitley 1997 or Ramaswamy-Sloek 1997).

A second question is that of comparing the degree of reaction of the various sectors of the European economies to the monetary impulses. Indeed, a macroeconomic monetary analysis masks the disparities of transmission in the sensitivity of the shocks of the different sectors of the economy as well as the differences in the degrees and the speed of reaction of these sectors within the economic cycle. This work fits into this problem. Our aim is to compare the sensitivity of European agricultural sectors to monetary policy impulses in order to specify the effects to be expected from the single monetary policy on these sectors. This work therefore lies at the intersection of analyzes of the sensitivity of European economies to monetary policy and analyzes of the sensitivity of the agricultural sector to the impulses of macroeconomic policy (see, for example, Orden-Fackler, 1989; Meyer, 1987; Kirrane 1993).

From a theoretical point of view as well as from an empirical point of view, the determination of the effects of monetary impulses on economic activity directly enters into the still largely open debate on the real effects of nominal pulses. However, it is possible to retain a general presentation and classification of the different transmission channels of the real effects of monetary policy (see, for example, Bernanke-Gertler 1995, Kirrane 1995 or Taylor, 1995). Monetary policy results in a distorted nominal interest rate structure that equates to a distortion of the structure of real rates under a short-medium term price rigidity assumption. The resulting change in the cost of capital is then reflected in the interest-sensitive components of aggregate demand. This effect can be considered as the direct effect of monetary policy. In flexible or semi-flexible exchange rate regimes (as in the European Monetary System), monetary policy also has an indirect effect on activity via exchange rate movements (Kirrane 1994). The international financial adjustments induced by the variation of the interest rates lead here to a modification of the nominal parity of the currency of the country considered. Under the assumption of rigidity has in short-term prices, this change in the nominal exchange rate is equivalent to a change in the real exchange rate that affects the components exported and imported from demand. These macroeconomic effects can be transposed to the sectoral level. However, if the assumption of supply exogeneity is acceptable at the macro-economic level because of the exogeneity of the available quantity of labor and capital factor in the short-medium term, it is difficult to retain this assumption at the macroeconomic level. Changes in interest rates and exchange rates are therefore reflected in both supply and sector demand. In addition, while the sector is "small" relative to the rest of the economy, sector demand is also affected by income variation at the macro level. Overall, it is possible to represent the main effects of monetary policy on sectoral activity from a model representing supply and demand behavior at the level of any sector.

The supply of the sector is represented by a maximization program carried out from domestic profit inputs in which the production the price depends on is the real interest rate, of imported inputs whose price depends on the exchange rate. and other inputs:

{MaxUn = Pn.Yn-Pz (r) .Z-Pze (e) .ZE-Pa.A [Y = F (Z, ZE, ZA)

Notations: Tnn benefit of sector n, Pn : output price of sector n, Yn: output of sector n, r: real interest rate, e: real exchange rate (one unit of foreign currency equals e units of domestic currency), F / r : price of domestic inputs sensitive to the real interest rate, Z: domestic input whose price is sensitive to the interest rate, Pu (e): prices of imported inputs sensitive to the real exchange rate, ZE: imported inputs of which the price is sensitive to the real exchange rate, Pa: price of other inputs, A: other inputs.

The Pzet Pu prices are assumed to be increasing functions, respectively of the real interest rate r and the real exchange rate e. Under the usual assumptions about the production function, the sector supply function n (Yns) derived from the profit maximization is given by:

YDn = YDn (Y, Pn / P) (2) _ft>* & ft> "A or

K = Ysn (Pn, Pz (r), Pze (e), Pa) (Dwith dYsn (.) / DPn> 0, dYsn (.) / DPz <0, dYsn (.) / DPze <0 ttdYsn (.) / dPa <0

The demand for the sector n is assumed to depend both on a traditional income effect which is here represented by the effect of aggregate national output (Y) and a relative price effect represented by the ratio between the output price of sector n and the general level of prices (P):

with dYDn (.) / dY> 0 and dY%) / d (Pn / P) <0

This work being focused on the sector In the agricultural sector, the output price of sector n is assumed to be independent of the real interest rate. Figure 1 shows the supply and demand functions 3 ft) u ft) M (A ft) sectorial corresponding to equations (1) and (2).

In the case where the sector is a competitive sector, the activity of the sector is determined by the intersection of the supply and demand curves. When support systems influence supply and / or demand, the level of activity in the sector is determined from the supply function and depends on the type of support and the level of support.

A decrease in the positive monetary interest rate impulse is caused in the sector a by a reduction in the price of interest-rate domestic inputs Z (Pz (r)) such as the capital input. In Figure 2, the result is an increase in sector demand until supply YSN derivative "of input. The Z feeds increase in the correlative level of macro, increasing the component of the aggregate demand sensitive the interest rate. The effect indirect monetary policy transiting the exchange rate will partially diminish this outcome. Indeed, the depreciation of the nominal exchange rate induced the reduction in interest rates increases the price of imported inputs PJe) and thereby reduces the optimal supply of sector n, this reduction in supply is for example by the displacement of the curve of 'offer YSN the net effect of these direct and indirect transmission channels of monetary policy on sector offers here depends on the relative size of these two effects.

On the demand side, the expansionary effect of the reduction of the rate of Interest is translated at the macro-economic level by an increase in income in the short-medium term, which benefits all sectors of the economy and sector n in particular. This demand expansion effect is eventually reinforced by a substitution effect in demand for domestic products due to the depreciation of the national currency, which increases the competitiveness of domestic products compared to imported foreign products. From a theoretical point of view, the impact of monetary policy on the activity of the sector is therefore a priori indeterminate (Kirrane 1994). However, an overall positive effect is conceivable when the share of imported inputs is low and the share of capital inputs is high.

In the case where the activity of the sector is regulated by a set of production and / or demand support systems, the net effect of these shifts of the supply curves on the sectoral activity depends on the support systems in force and the level of support.

The analysis of impulse transmission the activity of a sector and the agricultural sector in particular is generally based on VAR model This has the advantage of capturing all the dynamic effects of monetary shocks on sector variables while avoiding the choice of a particular structural form to model all the transmission channels of these shocks. In view of an international comparison of these transmission channels, this approach seems to be largely justified since it allows to retain a homogeneous and identical modeling for all countries while avoiding the thorny problem of modeling specificities. Despite its "black box" aspect, VAR modeling is also suitable for estimating statistical relationships between a set of variables without trying to detail the relative size of the different mechanisms at work (Sims, 1992, Kirrane 1994). We therefore retain here this methodology to evaluate the net effects of monetary impulses on the activity of the agricultural sector. In a standard VAR approach, a vector of relevant variables is modeled from the offsets of this vector. Here we are moving away from the usual approach of two-level VAR modeling.

First, we use a "partial" VAR model in which some variables are modeled by a specific set of variables. Since the agricultural sector is "small" at the macroeconomic level for almost all European countries2, we assume that the block of monetary and macroeconomic variables is not influenced in return from the agricultural sector. Moreover, as the agricultural sector is characterized by a set of specificities relative to the rest of the economy (climatic hazards, support systems, etc.), the modeling of the activity of the agricultural sector in each country systematically retains the activity European agricultural "partial" as finally variable retained explanatory. has the following form:

The model VAR POLM, = ClNF + oc oc21 k = 1 P a22VARMACt_k +2 oy V) JS k = 1

UPt0LM a (3) is VARMAQ = ClNF + k = £ I a2xPOLMt, k + P £ OL22VARMACt_k +

l / JWF (4) yA = CYA + * = 2 l <h \ POLMt_k + t = 2 l oc32 VA /? MAC _fc + t = i S <4 * A, _t + * £ = o a54rA £ [//? r_, + f / fA (5)

is still in the form of a standard VAR write:

at (L) oc12 (L) 0 0 a21 (L) a22 (L) 0 0 a31 (L) a32 (L) ( X33 (L) oc34 (L) POLM VARMACX YA, YAEURX c2 Pi

When this variable is not available over the entire period, we retain the money supply as defined in M2 (Netherlands, Finland, Greece, Sweden, Ireland). The agricultural production variable is the volume agricultural production series provided by FAO (volume production index, base 100 in 1965-1968). The data are annual and cover the period 1965-1995 for all countries with the exception of Italy (1966-1995) and Spain (1975-1995).

Finally, whose model equations VAR do not contain a model not the same "partial" regressors, the models are systematically estimated by Zellner's SUR method. Once estimated, the VAR (6) form is inverted to obtain the innovation response functions. The estimated VAR model is then considered as the reduced form of an underlying structural model of the type k = 1 In which the vector of structural shocks W satisfies E (UrUt) = X diagonal and is both identifiable and interpretable.

These structural shocks are related to the innovations estimated from the VAR model by the relation Ut = B ~ o, Wr II thus comes:

E (UrUt) = (7) and the identification of the structural shocks goes through the determination of the elements of the matrix Bo from equation (7). For this, we assume here that the lower triangular Boest matrix so as to apply a Cholesky decomposition. The chosen triangular structure imposes the following two identifying hypotheses: - changes in the nominal interest rate (or the money supply) have no instantaneous effect on the rate of inflation - changes in activity have no instant effects on the rate of inflation and on the interest rate. For the order each POLM-country, VARMAC- the variables YA and are the functions thus classified from in responses are obtained from the vector moving mean form:

Xt = = E + F (L) .Ut (8)

Several impulses studies 1992, empirical monetary Christiano-Eichenbaum-Evans adopt relative this to strategy estimation mechanism 1994) of transmission (Ber- (6) ci 52 ■ DS! with a /; (X) = k = ^ a /; y, = 1,2,3, ef a34 (X) = ^ a ^ L

Notations: POLM : representative variable (or vector of representative variables) of the monetary policy V ARM AC: macro-economic variable (or vector of macro-economic variables) YA: YAEUR: representative variable representative of the activity of the agricultural activity agricultural in Europe. The estimation of a VAR model is generally made from OLS on stationary variables (or made stationary) that provide efficient estimators. However, the fact of making stationary variables deletes any relationships long term between the levels of the variables Given the small number of observations available for each country (T = 30 for most countries), we prefer to estimate the VAR model directly on the levels of the variables particular long-term relationship not being suggested here.

Estimation and empirical results A VAR model similar to (6) is estimated for Belgium, France, Germany, Italy, Ireland, the Netherlands, Portugal, Finland, Spain, Greece , Sweden and Great Britain. The first nine countries participate in the Monetary Union on the basis of their situation with regard to the Maastricht convergence criteria.

For each of these countries, the VAR model incorporates a representative variable of monetary policy, the general price level and an indicator of agricultural production. Taking into account the general level of prices is justified by the fact that the two main transmission channels of monetary policy are the real interest rate and the real exchange rate. By conditioning the model by the general price level, it is therefore possible to identify the share of changes in the real exchange rate and the real interest rate induced by changes in the nominal exchange rate and the nominal interest rate.

The interest rate at 3 months is used as the representative variable for monetary policy for most countries (Belgium, Germany, France). The responses are normalized to correspond to a unitary shock. The response functions of the activity variable nominal agricultural interest rate at one impulse (or at a positive monetary unit impulse on the unit negative for some countries) are shown in Figures 3 (standard deviations are not provided for the sake of clarity in the graphs)

The importance of monetary impulses in the determination of agricultural activity is also assessed from the calculation of the decomposition of the variance of the forecast errors of agricultural production at q forward periods From the vector moving average form (8), the variance of the forecast error at q periods of the variable YA (VEP (YA, q) ) and the share of this variance due to the innovations of the monetary policy variable (VEP (YA, POLM, q)) are respectively given by:

q 3 VEP (YA, q) = X k = 0l = £ (F * y / M) 1 VEP (YA, POLM, q) =

The relative importance of the variable POLM in the explanation of the variable YA on the horizon q (PARTVEP (YA, POLM, q)) is thus defined as the proportion of the variance of the prediction error q periods ahead of the POLM variable that can be explained by the innovations of the POLM variable: PARTVEP (YA, POLM, q) = VEP (YA, POLM, q) / VEP (YA, q) Table 1 presents the estimated values PAR-TVEP (YA, POLM, q) for the selected countries and for a number of periods ranging from 1 to 20 years.

The strongest effects appear in Spain, Ireland and a lower level in Greece and in the Netherlands. In all other countries the share of the variance of the forecast error at 15-20 years induced by the monetary policy shocks is of the order of 9 to 12%. This criterion thus seems to reveal a core of countries characterized by a sensitivity of agricultural activity to relatively limited and homogeneous monetary shocks.

The joint examination of the response functions and the decomposition of the variance of the forecast error shows that the increase in interest rates or the introduction of a restrictive monetary policy is accompanied by a reduction of the activity of the agricultural sector in all the selected European countries. The only "atypical" answers concern Germany and Sweden, where agricultural activity begins to shrink and then increases again. These "atypical" reaction functions can largely be explained by the simplicity of the selected models.

In most "typical" cases, the impact of rising interest rates on agricultural activity is limited and is taking place relatively slowly. In comparison with the results obtained by Ramaswamy-Sloek (1997), Barran-Coudert-Mojon (1996) or Gerlach-Smets (1995), the multiplier effects of monetary policy on the activity of the agricultural sector are both lower and slower to appear than at the macro level. Indeed, the multiplier effects of monetary policy on GDP are fully realized at the end of the sixth quarter following the shock for the "fastest" countries and at the end of the tenth quarter for the other countries. The response functions here reveal that the effects on the agricultural sector are only realized at the end of the sixth or seventh year for the "fastest" countries.

The comparison of the results obtained for the various countries shows that these countries are relatively different, on the one hand, in terms of the impact of monetary policy on agricultural activity and, on the other hand, in terms of the speed of reaction of the agricultural activity.

Countries can be ranked according to their rate of reaction to monetary policy and the size of the associated multiplier effects by calculating for each country the maximum value of the cumulated multiplier effect (expressed in absolute value) and the number of years required for 90% of this multiplier effect to be achieved. The effects of the interest rate on agricultural activity are both the weakest and the fastest to take hold in Great Britain and Germany, where almost all the effects are felt before the fourth year in a row. Belgium, Portugal and France are in an average position both in terms of the size of the multiplier effects and the speed of reaction of agricultural activity to interest rate shocks. Almost all of the multiplier effects are realized between the fifth and the tenth year. Italy is characterized by a multiplier effect of the interest rate comparable to those almost of this two group times more slow country but the speed reaction rate reaction is of the agricultural activity of Spain is comparable to that of Italy but the multiplier effect is clearly almost six times stronger than in the Italian case and almost twice stronger than in France.

Multiplier effects induced by an interest rate of shock cannot be directly compared to the size of the effects induced by a monetary shock. Such a comparison requires the preliminary estimation of the elasticity of the interest rates to the money supply, but a "means" classification and "strong" of the effects in each multiplier of the cases allows in "weak", to characterize the response of the different countries to a common monetary shock. he countries Germany that respond and Britain the faster and the less are at the strongly to this impulse. Belgium, Portugal, France, Greece, Finland and the Netherlands have both a reaction rate and an average shock sensitivity, the impact on Italy is similar but the speed of reaction in this country is much slower. the most important and the slowest are Spanish In addition to these two countries, the effects of monetary policy via the interest rate channel are relatively limited in the agricultural sectors of European countries. The introduction of a common monetary policy should therefore not create significant disparities in the operating conditions of the agricultural sectors of European countries. This conclusion, however, would need to be reconsidered from an analysis based on a finer modeling of the agricultural and monetary specificities of the different countries. Similarly, the use of a theoretical intertemporal formalization would capture the effects of the interest rate on the valuation of the heritage values updated of the agricultural sector and on the induced wealth effects.

Conclusion

Assessing the effects of the common monetary policy on European agricultural sectors involves modeling the effects of interest rates on these sectors. We use an approach in terms of a "partial" VAR model to evaluate both the multiplier effects of a monetary policy shock on the agricultural activity of European countries and their speed of implementation.

European countries appear very different from the point of view of these two criteria. However, the effects of interest rates on agricultural sectors are both limited in size and relatively slow to manifest. In addition, the multiplier effects of monetary policy on the activity of the agricultural sector appear at first weaker and slower than the effects of monetary policy on macroeconomic indicators of activity such as real GDP.

The agricultural sectors most sensitive to monetary impulses are those of Spain and Ireland. A relatively large country core (Belgium, Portugal, France, Greece, Finland and the Netherlands), however, seems to be characterized by a "medium" sensitivity to the interest rate and a moderately slow reaction rate of around 7 to 10 years.

BIBLIOGRAPHIC REFERENCES